# Multi-charge Transfer from Photodoped ITO Nanocrystals

Michele Ghini,[1,2] Andrea Rubino,[3] Andrea Camellini,[3] Ilka Kriegel[3,a]

[1] *Department of Nanochemistry, Istituto Italiano di Tecnologia, via Morego 30, 16163 Genova, Italy*

[2] *Dipartimento di Chimica e Chimica Industriale, Università degli Studi di Genova, Via Dodecaneso 31, 16146 Genova, Italy*

[3] *Functional Nanosystems, Istituto Italiano di Tecnologia (IIT), via Morego 30, 16163 Genova, Italy*

a Author to whom correspondence should be addressed: Ilka.Kriegel@iit.it

Metal oxide nanocrystals are emerging as extremely versatile material for addressing many of the current challenging demands of energy-conversion technology. Being able to exploit their full potential is not only an advantage but also a scientific and economic ambition for a more sustainable energy technology. In this direction, photodoping of metal oxide nanocrystals is a very notable process that allows accumulating multiple charge carriers per nanocrystals after light absorption. The reactivity of the photodoped electrons is currently the subject of an intense study. In this context, the possibility to efficiently extract the stored electrons could be beneficial for numerous processes, from photoconversion and sunlight energy storage, to photocatalysis and photoelectrochemistry. In this work we provide, via oxidative titration and optical spectroscopy, evidence for multi-electron transfer processes from photodoped $Sn:In_2O_3$ nanocrystals to a widely employed organic electron acceptor (F4TCNQ). The results of this study disclose the potential of photodoped electrons to drive chemical reactions involving more than one electron at a time.

## I. INTRODUCTION.

Metal oxide (MO) semiconductors are inorganic materials of great interest in the field of optoelectronic applications for energy-related technology.[1] They offer an ideal combination of several properties ranging from environmental stability, chemical tunability[2] to optical transparency and good charge mobility.[3] Recently, the ability to modulate their charge carrier density through aliovalent substitutional doping and post-synthesis methods is boosting the implementation of doped MO nanocrystals (NCs), such as Sn-doped $In_2O_3$ (Indium Tin Oxide [ITO]) NCs.[4–7] Doped MO NCs, in fact, combine both the ability to efficiently harvest NIR sunlight radiation through the excitation of

doping- and size- dependent localized surface plasmon resonances (LSPR) and the potential to store multiple delocalized photo-excited carriers.[8–10]

Photodoping, a light-driven charge accumulation of electrons induced by multiple absorption events of high-energy photons (beyond the bandgap of MO semiconductor), is emerging as a contactless and promising tool to promote the photo-conversion process.[7] Photodoping of ITO NCs in presence of a sacrificial hole scavenger, such as ethanol, under anaerobic conditions, results in the reversible accumulation of tens to hundreds of electrons per single NC.[11] The reversibility of the photodoping process (and therefore the removal of photo-chemically excited electrons) was initially demonstrated via potentiometric titration with the addition of a molecular oxidant to the colloidal suspension of MO NCs.[12] Due to the proportionality between the plasmonic features and the free carrier density $n_e$,

$$\omega_{LSPR} \propto n_e^{\frac{1}{2}}, \text{ and } \sigma_{Abs} \propto n_e^{\frac{2}{3}},$$

being $\omega_{LSPR}$ and $\sigma_{Abs}$ the central LSPR position and its peak absorption cross section, [13] the storage and depletion of photodoped electrons upon photodoping and oxidative titration can also be performed by monitoring the spectral evolution of the LSPR.[7,14] In this context, the opportunity to access multiple charge transfers events of photodoped electrons would be enormously relevant in terms of efficiency, not only for photovoltaic energy conversion purposes, but also in applications such as batteries, or implementations in photoelectrochemistry and photocatalysis.[15–17] The transfer of more than one electron per unit of active component (i.e. MO NCs) implies an increase in energy density which also translates into a reduced amount of materials to be employed and lower costs. This energetic and economic principle, perfectly in line also with current sustainability policies, is obviously valid for the redox transformations of electrochemical cells and for catalytic reactions.[18] Whether the transfer takes place in solution or in solid systems such as electrodes, the participation of multiple charges can enhance the performance in terms of reactions yield and kinetics. Such beneficial process has already been observed in many reports, as in the case of gold nanoparticles for $CO_2$ conversion or organic donor-acceptor systems capable of proton-coupled multi electron transfer.[17,19,20] Similarly, colloidal semiconductor nanocrystals demonstrated of being able to release multiple electrons, after the absorption of light, which can then interact with organic compounds or be suitably stored for the subsequent production of current.[21] Until now, only few MO related studies have specifically addressed the possibility for multiple electron transfer of photo-excited electrons. For instance, ZnO and $TiO_2$ NCs, have found successful use in electrochemical applications such as oxygen reduction reactions, which typically involve multiple electron transfer processes. [22–25]

In this work, via oxidative titration and optical spectroscopy, we investigate the ITO NCs ability to provide multi-transfers of electrons accumulated after the photodoping process. Specifically, to address multi-electron oxidation we made use of a chemical compound capable of undergoing a double ionization, namely, 2,3,5,6-tetrafluoro-7,7,8,8-tetracyanoquinodimethane (F4TCNQ). The concentration dependent evolution of the F4TCNQ ionized species, with their peculiar spectroscopic features, will help in discerning the kinetic of charge transfer as a stepwise sequence of concerted phenomena.

## II. METHODS.

For the synthesis of nanocrystals, we adopted the continuous growth mechanism starting from a solution of the precursors, indium (III) acetate (Sigma-Aldrich) and tin (IV) acetate (Sigma-Aldrich) in a 1:9 = Sn:In ratio. The two acetate were left in oleic acid (Sigma-Aldrich) at 150°C under $N_2$ for several hours. The precursors were then slowly injected (at a rate of 0.30 mL/min) with a syringe pump into 13mL of oleyl alcohol at 290°C. The solution was then washed (twice) with ethanol (~12mL) and centrifuged at 7300 rpm for 10 mins. Finally, the colloidal nanoparticles were dispersed in hexane (Sigma-Aldrich) as a stock solution with a concentration of 13 mg/mL. F4TCNQ titrant were prepared by dissolving 0.34 mg in 40 mL of anhydrous toluene (Sigma-Aldrich). In order to avoid any contact with external oxygen, titrant addition steps were carried out in the inert environment of an argon filled glovebox. For both photodoping and titration experiments, rectangular anaerobic cuvettes with a sealed screw cap, optical path of 5 mm and volume of 1.4 mL were used (Starna Scientific). The photodoping process was carried out by illuminating the cuvette containing the solution of ITO NCs dissolved in anhydrous toluene with a UV LED (central wavelength: 300nm, bandwidth: 20 nm) placed at a distance of 12 mm from the cuvette window (Thorlabs M300L4). UV power density at the front window of the cuvette was 36.8 mW $cm^{-2}$. Monitoring of the optical response of both as-prepared ITO NCs upon photodoping and ITO/F4TCNQ mixture after each titration steps was carried out by measuring the Absorbance spectrum using a UV-Vis-NIR spectrophotometer (Agilent Cary 5000). Quantitative and structural characterization of nanoparticles were carried out via inductively coupled plasma mass spectrometry (ICP-OES) and transmission electron microscopy (JEOL JEM-1400Plus - 120 kV TEM/STEM).

## III. RESULTS AND DISCUSSION

### A. ITO NCs photodoping.

The ITO NCs analyzed in this work have diameter of about 11 nm (**Figure 1a, b**) and are colloidal particles stabilized by organic ligands. More precisely, the organic part consists of oleate molecules. In view of the study on electro optic behavior through the titration, we took an aliquot of the colloidal solution in hexane, let the nanoparticles dry and we transferred them into a glovebox where we prepared a solution in toluene anhydrous (~ 0.1 $10^{-9}$ mol/L). The solvent exchange is dictated by the low solubility of F4TCNQ molecules, employed as a diagnostic tool for assessing the electron transfer, and because inert atmosphere conditions are strictly needed for the photodoping process occurring in a liquid environment.

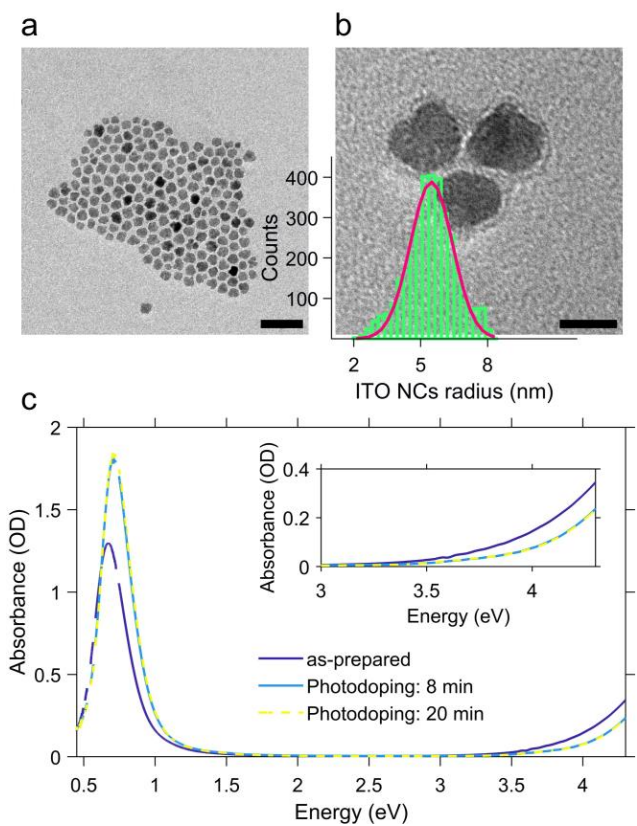

**Figure 1** Transmission electron microscopy images of as prepared Sn-doped $In_2O_3$ (ITO) nanocrystals, scale bar 50 nm (a), 10 nm (b). Panel (b) also shows the nanocrystals' radius size distribution obtained via statistical analysis of the transmission electron microscopy images. Absorbance spectra of as prepared and photodoped ITO nanocrystals (c) dissolved in anhydrous toluene. Photodoping is performed by illuminated the ITO nanocrystal solution with a UV LED (central wavelength: 300nm) with increasing exposure time with respect to as prepared conditions (8, 20 minutes). Inset of panel c, highlight the effect of photodoping on the band edge absorption (Burnstein-Moss effect).

In **Figure 1c** we report the results of the photodoping experiment on ITO nanocrystals. As can be seen from the absorption spectra recorded upon a gradual increase of illumination times, the UV excitation alters the optical response of ITO NCs. In particular, spectral changes affect both the plasmonic peak and the bandgap. As the exposure time is prolonged, the LSPR feature associated with the plasmonic absorption increases in intensity and moves towards higher energies. Similarly, the bandgap onset also undergoes a blue shift. The effect on these absorption properties reaches saturation after 8 minutes of UV exposure. At this point, we want to underline a substantial difference with respect to the usual conditions adopted in the doping treatment with ultraviolet light. In order to keep both the overall charge neutrality and stabilize the accumulated electron density of each nanocrystal, the intervention of a hole-quencher (e.g. ethanol, methanol) is necessary. In our case, however, the increase in the density of charges and the consequent increase in the plasmonic signal is quite evident without the addition of further compounds to the solution in toluene. Such behavior indicates that the recombination of the photogenerated charges is most likely suppressed thanks to the presence of the ligands and/or surface states as already demonstrated in other cases in the literature.[26–28] The bleaching of the band edge absorption is the result of the widening of the bandgap (Burstein-Moss effect[29]). This effect also reported for different MO NCs (e.g. doped indium oxide) further confirms the increase of electron density upon photodoping.

### B. ITO/F4TCNQ Titration.

Once the maximum intensity of the LSPR peak reached saturation, we started the quantitative analysis of the number of electrons released by the photodoped ITO NCs by successive addition of oxidative titrants. As already mentioned, in this study, we selected F4TCNQ, a well-known electron acceptor with a high value of electron affinity (5.2 eV), widely employed both for fundamental studies of charge transfer processes (e.g. photoluminescence quenching) and as a p-dopant of organic materials in hole-transporting layers for photovoltaic applications.[30–32] More specifically, the characteristic that makes this compound particularly interesting in the analysis of multiple electron transfer is its ability to acquire, in its neutral form, up to two electrons. F4TCNQ is thus able to form two distinct ionized species[33] with donor-host environment having lower values of ionization energy ($\lesssim$ 5eV). Moreover, the formation of stable anion and dianion species lead to different spectral features in the optical absorption of the donor-F4TCNQ mixture therefore allowing the possibility to distinguish between single and double ionization process with routinely available optical measurements. The formation of F4TCNQ dianion is observed in several polymer-dopant systems.[34–36] Recently, doping of bithiophene–thienothiophene-based copolymers with F4TCNQ showed an almost complete double ionization of dopant molecules providing a ionization efficiency of 200% which also resulted in an enhancement of charge carrier transport.[34,37] To the best of our knowledge double reduction of F4TCNQ has never been reported neither for inorganic-dopant systems nor for photodoped MO NCs-dopant mixtures.

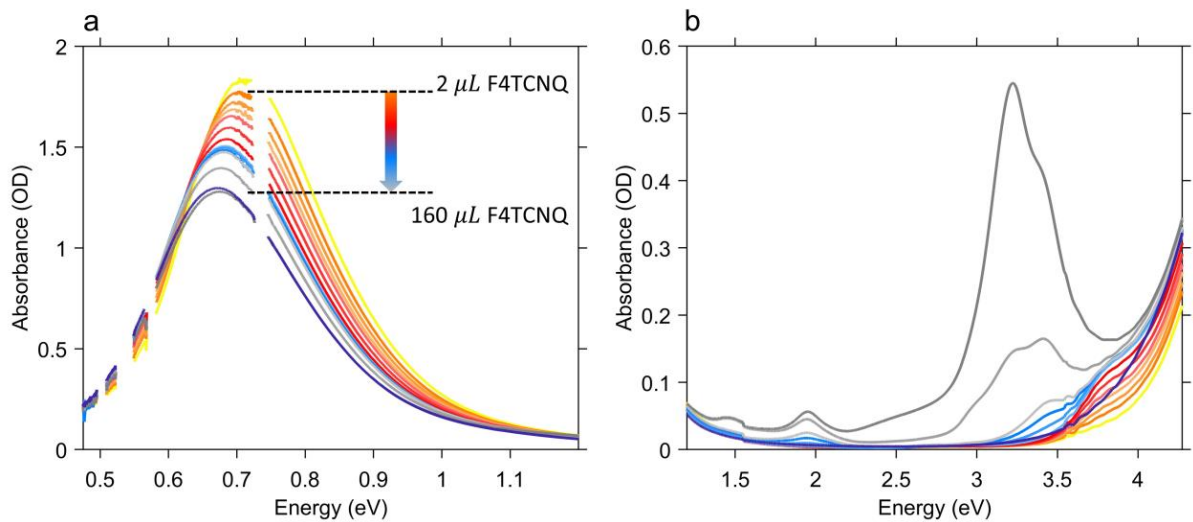

**Figure 2** Results of the oxidative titration of photodoped Sn-doped In$_2$O$_3$ (ITO) nanocrystals. Panel (a) shows the recovery of the localized surface plasmon feature associated with plasmonic absorption upon adding increasing amount of F4TCNQ molecules. Panel (b) shows the titration effects induced on the ITO bandgap absorption. Dark blue and yellow lines in panels (a) and (b) correspond to the as-prepared and photodoped (exposure time: 20 minutes) ITO NCs absorption spectra. During titration, Absorption spectra were recorded after adding a total sum of 2, 4, 6, 8, 11, 14, 17, 20, 23, 26, 60 and 160 $\mu$L of F4TCNQ to the photodoped ITO NCs solution.

**Figure 2** shows the results of the titration experiment on ITO NCs. In order to accurately establish the number of electrons extracted from the ITO NCs we conducted titration by gradually adding an increasing amount of F4TCNQ molecules from a 0.3 mM stock solution and monitoring over time through absorption, the formation of reduced F4TCNQ species (molar ratio between 88 and 98 mol%, see Table S1 in Supplementary Material file). Figure 2a displays the effects of the addition of the electron acceptor on the LSPR peak of the ITO NCs in the NIR range. In this case, we observe that through titration the system re-establishes its as prepared conditions: the LSPR peak decreases in intensity and undergoes a red shift with increasing amount of F4TCNQ. Both are signatures of the reduction of charge density in ITO NCs , which have reacted with the F4TCNQ molecules. The photogenerated and charged-compensated electrons in the post-synthetic photodoping treatment are more reactive with respect to the electrons present due to the aliovalent doping after the incorporation of tin elements in the indium oxide, even in mild oxidation conditions.[11,38] Taking into consideration, instead, the spectral window in the UV-Vis range, as shown in Figure 2b, we can see a second effect on the absorption of the ITO. In conjunction with the recovery of the LSPR signal, the absorption band edge also tends to return to the pre-photodoping spectral shape by compensating the blue shift originated by the Burnstein-Moss effect. Interestingly, in this region, the appearance of a second contribution around 3.7 eV is evident. This contribution to the overall absorption recovery increases in its intensity only during the first titrant additions (orange and red spectra in figure 2b) and, as detailed below, represents to the formation of the F4TCNQ dianion in the ITO NCs/F4TCNQ mixture. Moreover, in the latest addition of titrants (> 100 µL), dilution effects start to play a role in the decrease of the LSPR peak intensity. At this stage, is therefore convenient to switch the focus towards the information obtained from the analysis of the neutral and ionized species, following systematically the titration process.

**C. Multi electron transfer**

As previously discussed, the aim of titration in this work is twofold: to quantify the amount of charges that can be extracted per photodoped nanocrystal and to analyze the electron transfer process from a kinetic point of view. For this purpose, we want to briefly recall the established spectral features associated with F4TCNQ species and their photo-physics. The absorption peak of the neutral molecule has a maximum around 3.2 eV whereas the peak of its anion shows a maximum around 3 eV and two characteristic peaks between 1.3 and 2 eV. The double ionized molecule (dianion) present an absorption peak around 3.7-3.8 eV. In addition, this compound is able to react through integer charge transfer and / or form charge transfer complexes.[39,40] The ionized species of F4TCNQ are also particularly reactive and can interact with each other or even dimerize.[41,42] All these transformations can generate new optically allowed transitions that "intervene" in the absorption with new peaks that may appear along with those already mentioned.

Hence, in the titration process, the absorption spectra recorded from the solution of photodoped ITO NCs/F4TCNQ molecules contains the contributions from all the species and charge transfer states that can be formed by reduction with electrons accumulated in ITO NCs. The spectra shown in Figure 2b are in fact a convolution of several contributions. As already observed in Figure 2, the superposition of the signals coming from the oxidant and the reductant concerns the UV-Vis region. Clearly, the absorption measurements recorded right after the addition of F4TCNQ to the solution of photodoped ITO NCs revealed the presence of the dianion alone among the F4TCNQ species (see Figure S1 in Supplementary Material file). In order to correctly analyze the behavior of the titrant and separate it from the response of the ITO NCs we fitted the experimental absorption spectra in the UV range with a polynomial function, which takes into account the ITO bandgap recovery in UV range upon titration (see Figure S2 in the Supplementary Material). **Figure 3a** displays the resulting peak obtained by subtracting this latter effect. The characteristic F4TCNQ dianion peak appears and increases in intensity with subsequent additions up to 14 µL of titrant solution (Figure 3b). This result demonstrates how photodoped ITO NCs are capable of a two-electron transfer to directly double ionize the F4TCNQ. This is the first evidence of such optoelectronic response from ITO NCs.

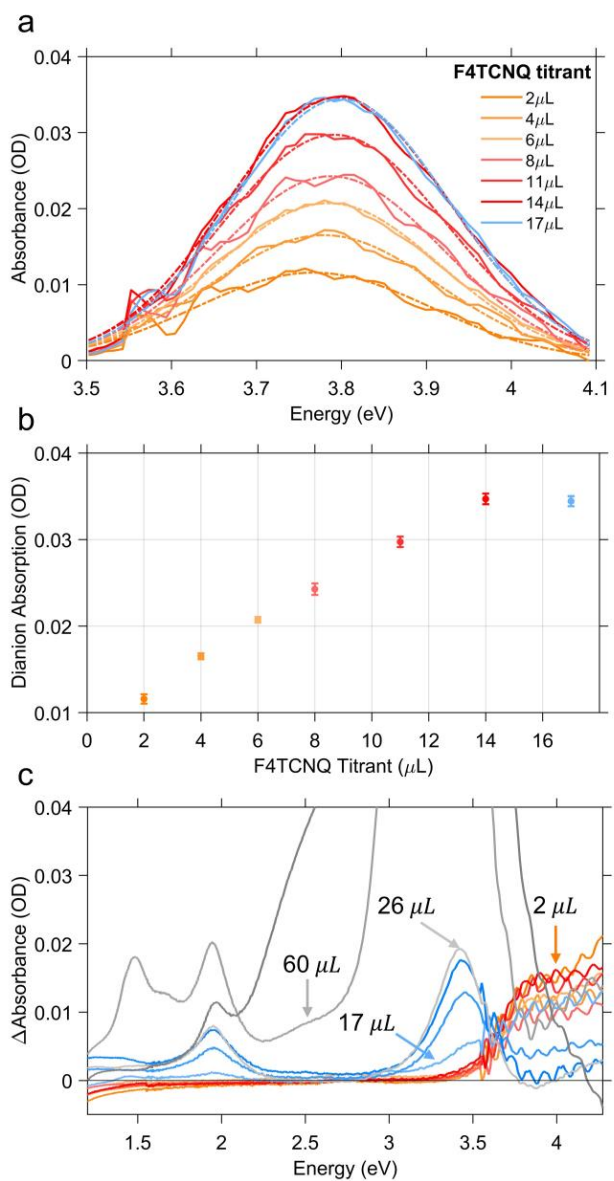

**Figure 3** F4TCNQ dianion contribution to the absorption spectra of ITO NCs/F4TCNQ mixture. Dash-dotted line are obtained by fitting dianion contribution with a Gaussian function (a). Dianion contribution increases linearly until the addition of $14 \mu L$ of F4TCNQ (b). Panel c represents the incremental effect of titration on the absorption spectra of ITO NCs/F4TCNQ mixture. Differential absorbance $\Delta Absorbance$ is obtained by subtracting the spectrum from the previous titration step from each curve.

Continuing with the titrant additions, the double reduction effect reaches a saturation level. Figure 3c shows the sequence of spectra of the titration experiments. In this case, the spectrum from the previous titration step was subtracted from each curve. In this way, we wanted to highlight the change in F4TCNQ ionization regime. In fact, after the addition of 14 $\mu$L the formation of the neutral titrant species becomes visible. This suggests the interruption not only of the transfer of two electrons, but of the ionization in general. Actually, a contribution relative to the anionic species is also observed in the visible region, which increases, albeit slightly. This second observation could also indicate the possibility that the transfer of photogenerated electrons could continue with single transfers. As already mentioned, however, the F4TCNQ can incur different types of reactions, including back donations and dimerization as suggested by an isosbestic point at 3.6 eV (see Figure 3c) between the region of neutral and dianion absorption. For this reason, we refrain from interpreting this last stage of the titration in terms on number of electron extracted. In this step, moreover, even minimal dilution effects can come into play and affect the overall optical response of the mixture.

As a counter-proof of the behavior analyzed so far actually deriving from the interactions between the electrons accumulated through photodoping and the electron acceptor molecules, we also carried out a control experiment on a similar ITO NCs sample, adding the titrant molecules before the photodoping. The result, shown in Figure S3 of Supplementary Material, indicates that adding F4TCNQ to the solution containing ITO NCs does not generate any ionization and we can observe just the neutral-related peak increasing in intensity according to the added volume of F4TCNQ solution. Electrons added through tin doping are more stable and do not participate in the titration. Surprisingly, illuminating with the LED the mixture containing F4TCNQ and ITO NCs we directly observed the formation of the dianionic species without any relevant changes in the ITO plasmonic peak. This behavior supposes a substantial difference between the kinetics of photodoping and the reactivity of the photodoped electrons, in favor of the latter. However, in this case, we cannot properly analyze the result considering the fact that prolonged illumination with ultraviolet light not only contributes to photodoping but can also initiate a degradation process of the organic titrant. In any case, what is also evident is that as the neutral peak disappears, either because of the dianion formation or because of the photodegradation, the ITO NCs photodoping signal kicks in. The plasmonic peak increases since no more neutral titrant molecules are available for the oxidation.

Finally, considering the starting concentration of the ITO nanoparticles, the concentration of the F4TCNQ molecules added until the appearance of the neutral peak and the transfer of two electrons per F4TCNQ molecule, we were able to estimate the number of electrons extracted per ITO NC (more details on the calculation can be found in SI). The result obtained reveals a transfer of ∼123 electrons per nanocrystal. Given the complexity of the final titration phase, it is also possible that this number could be an underestimate of the real number. In any case, the value obtained is an indication of the enormous potential of photodoped ITO NCs, as a multi-electron extraction platform. This direct evidence certainly deserves further study, taking into account several factors: first of all, the optimization of the system both in terms of materials (size, shape and level of aliovalent doping), operating conditions of the photodoping procedure and lastly the stabilization of the charges with suitable hole scavengers.[23] The potential to enhance energy density through multi-charge transfer processes makes these systems extremely competitive in the current research market for energy materials.

## IV. CONCLUSIONS

In summary, in this work we investigated the possibility of using photodoped electrons of ITO NCs for multiple charge transfer processes via titration with F4TCNQ molecules. Being, the latter, a compound capable of a single and/or double ionization, we were able to clearly distinguish the kinetics of the reduction process. The result we report revealed several strengths of the nanocrystal-based system. The synthesized nanoparticles are able to store new electrons after the ultraviolet illumination without the need for additional hole-scavengers. The photodoped electrons are then able to ionize the F4TCNQ molecules with a concerted transfer of two electrons thus showing the potential of these highly reactive electrons to be employed as efficient multi-electron photocatalysts. The photodoped ITO NCs can release more than 100 electrons per NC, demonstrating strong competitiveness, for example in the new generation of high-density batteries. The photodoping process provides a non-invasive method and the nanocrystals retain their stability, making them attractive candidates for the efficient development of light-driven energy conversion/storage applications and for multi-electron photocatalysis.

## SUPPLEMENTARY MATERIAL

See Supplementary Material for more details on ITO NCs photodoping (raw data), the Polynomial fitting, the "inverted" photodoping experiment and the procedure for the extraction of the number of electrons from each ITO NC.

## Author Contributions

The manuscript was written through contributions of all authors. All authors have given approval to the final version of the manuscript.

**ACKNOWLEDGMENT**

For this work, the authors acknowledge the support of both European Union's Horizon 2020 European Research Council, under grant agreement no. 850875 (Light-DYNAMO), and European Union's Horizon 2020 Research and Innovation program under grant agreement no. 101017821 (LIGHT-CAP).

Data available on request from the authors: The data that support the findings of this study are available from the corresponding author upon reasonable request.

# Multi-charge Transfer from Photodoped ITO Nanocrystals

Michele Ghini,[1,2] Andrea Rubino,[3] Andrea Camellini,[3] Ilka Kriegel[3,a)]

[1] *Department of Nanochemistry, Istituto Italiano di Tecnologia, via Morego 30, 16163 Genova, Italy*

[2] *Dipartimento di Chimica e Chimica Industriale, Università degli Studi di Genova, Via Dodecaneso 31, 16146 Genova, Italy*

[3] *Functional Nanosystems, Istituto Italiano di Tecnologia (IIT), via Morego 30, 16163 Genova, Italy*

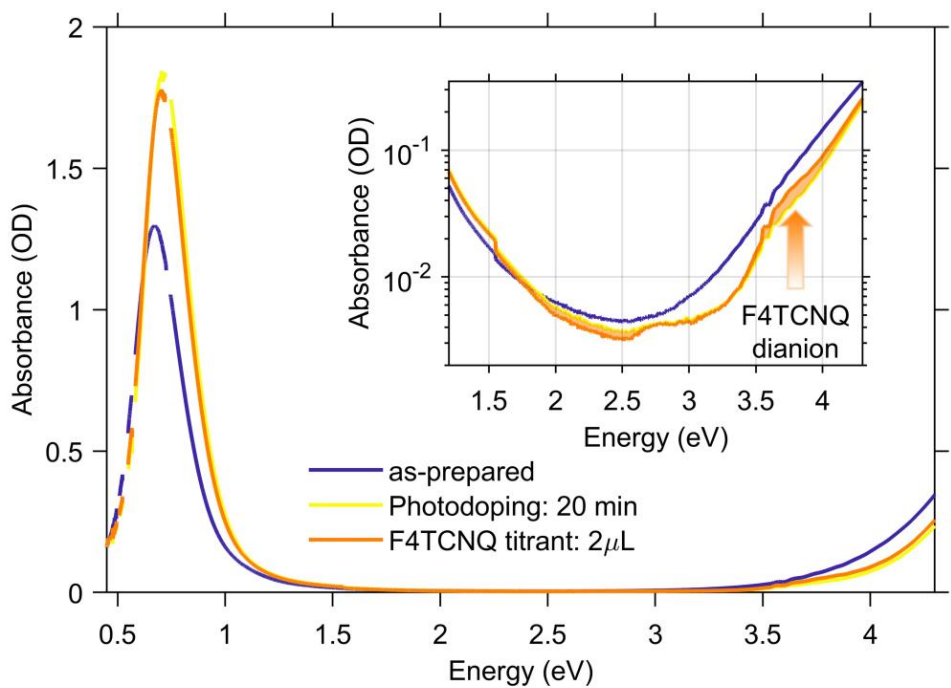

**Figure S4** Absorbance spectra of as prepared ITO NCs, photodoped ITO NCs (UV exposure time 20min) and ITO NCs/F4TCNQ mixture after the addition of the first titrant aliquot (2 $\mu$L, 88 mol%). Orange shaded area in the inset highlight the effect of the first titration step with respect to the photodoped ITO NCs solution. Only a small contribution between 3.5 and 4eV can be distinguished.

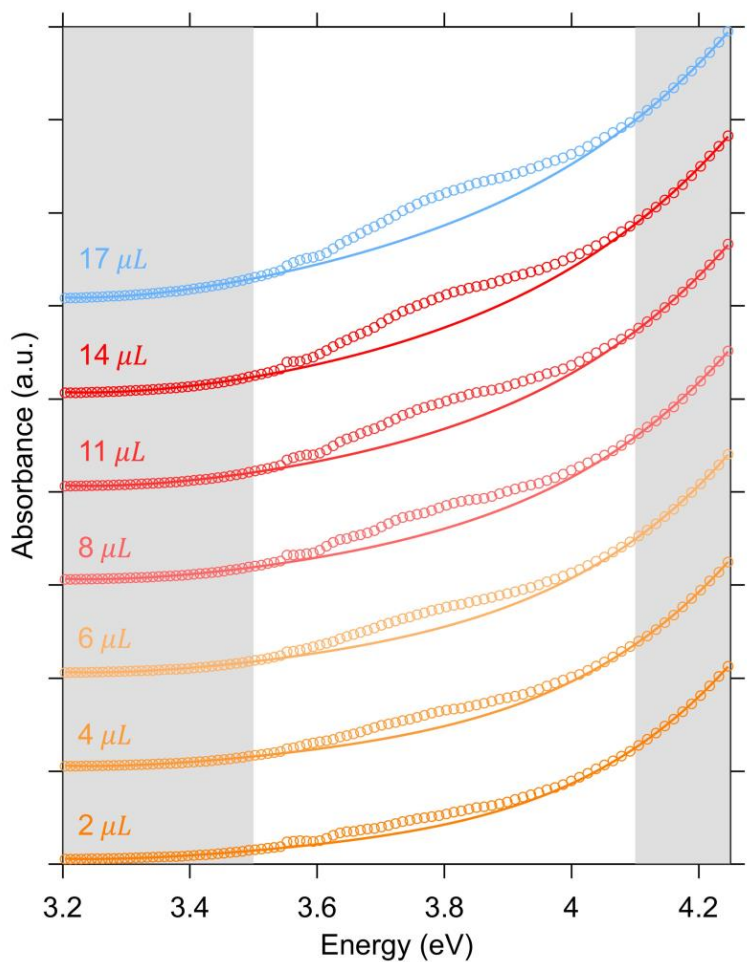

**Figure S5** Absorption spectra of photodoped ITO NCs after the addition of increasing amount of F4TCNQ titrant solution (0.3mM in anhydrous Toluene). In order to isolate the absorption of the F4TCNQ dianion from the recovery of the band edge we performed a 4$^{th}$ order fit of the absorption spectra. Fitting was performed by taking into account only those data points outside the region of dianion absorption (grey shaded areas).

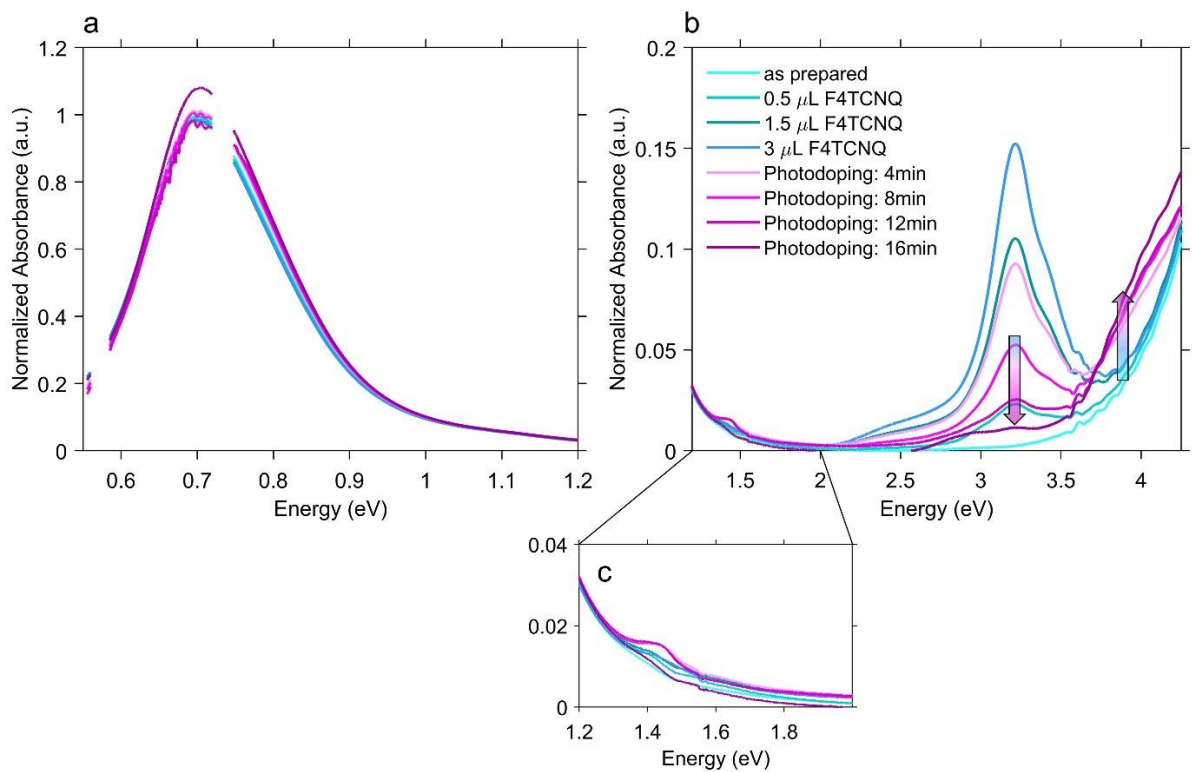

**Figure S6** Normalized absorbance spectra of as prepared ITO NCs, ITO NCs/F4TCNQ mixture for increasing amounts of F4TCNQ (0.5, 1.5, 3 $\mu L$) before any photodoping process. Absorbance spectra in magenta color scale show the effect of photodoping on ITO NCs/F4TCNQ mixture after the addition of $3\mu L$ of F4TCNQ. Panel (a) and panels (b, c) show the evolution of the spectra in the plasmonic and band gap regions, respectively. Arrows in Panel (b) display the effect of photodoping on the ITO NCs/F4TCNQ mixture. Panel (c) show a zoom in the spectral region of F4TCNQ anion absorption.

## Number of electrons extracted via Oxidative Titration of photodoped ITO NCs

We describe here in details the procedure followed to estimate the number of electrons extracted from each photodoped ITO NC upon oxidative titration. Given the concentration of titrant solution ($c = 0.085$ mg/mL) and the molecular mass of F4TCNQ molecules (276.15 g/mol), we can calculate the amount of moles of titrant molecules ($n_{F4TCNQ}$) added in the cuvette at each step of the experiment, as a function of the injected volume: $n_{F4TCNQ}(V) = cV/276.15$ (see Table S1). On the other hand, we calculated the weight of the average ITO NC from the volume of the unit cell (1.0355 nm$^3$), the weight of Sn, In and O atoms and the average radius of the NCs from TEM images (see Fig1b). In addition, from ICP-OES mass spectroscopy data we estimated the concentration of NCs present in the solution in order to extrapolate the number of ITO NCs moles ($n_{NC}$). Then, by normalizing $n_{F4TCNQ}$ with the moles of NCs ($n_{NC}$) present in the colloidal solution, we obtain the number of oxidizing molecule that reacted with each NC: $n_{reacted}(V) = n_{F4TCNQ}(V)/n_{NC}$. Finally, in order to estimate the number extracted photo-electrons for each NC, we multiply $n_{reacted}$ by a factor two or one, depending on the kind of reaction taking place (identified from dianion or anion peaks, respectively).

Thanks to the spectroscopic analysis showed in **Fig. 3**, we can safely assume that up to $V_1 = 14$ µL (volume of added F4TCNQ titrants) only dianion reactions occur, since no signatures of neutral nor anion peaks are present. We noticed that the next addition, $V_2 = 17$ µL, acts as a threshold value in the evolution of the reaction. At $V_2$, anion and neutral peaks start to appear in the UV-Vis range, and progressively increase with the further additions of titrants. After $V_2$, even if the LSPR peak continues to decrease and red-shift, suggesting that further extraction of stored electrons is still possible, the non-complete reaction of F4TCNQ molecules makes the analysis too complex to be carried on. Here, we counted two electrons for each F4TCNQ molecule which reacted until the midpoint between $V_1$ and $V_2$ ($Vmid = \frac{V1+V2}{2} = 15.5$ µL). Moreover, since the appearance of anion peaks indicates that the carrier extraction process still occurs at least up to $V_2$, we then added one extra electron for each molecule between $Vmid$ and $V_2$. Therefore the total number of electron extracted from each ITO NCs can be estimated by: $2e^- \times n_{reacted}(V_{mid}) + 1e^- \times (n_{reacted}(V_2) - n_{reacted}(V_{mid})) \approx 123$.

|  | Moles of ITO NCs ($n_{NC}$) [$10^{-9}$ mol] | | |
|---|---|---|---|
|  | 0.08 | | |
| **Added Volume of F4TCNQ [$\mu$L]** | **Moles of F4TCNQ ($n_{F4TCNQ}$) [$10^{-9}$ mol]** | **mol %** | $n_{reacted} = n_{F4TCNQ}/n_{NC}$ |
| 2 | 0.62 | 88.3 | 7.6 |
| 4 | 1.2 | 93.8 | 15.2 |
| 6 | 1.8 | 95.8 | 22.7 |
| 8 | 2.5 | 96.8 | 30.3 |
| 11 | 3.4 | 97.7 | 41.7 |
| 14 | 4.3 | 98.2 | 53.1 |
| $Vmid = 15.5$ | 4.8 | 98.3 | 58.8 |
| 17 | 5.2 | 98.5 | 64.4 |
| 20 | 6.2 | 98.7 | 75.8 |
| 23 | 7.1 | 98.9 | 87.2 |
| 26 | 8.0 | 99 | 98.6 |
| 60 | 18 | 99.6 | 227.4 |
| 160 | 49 | 99.8 | 606.5 |

**Table S1** Amounts of F4TCNQ titrants in terms of microliters and number of moles added to the photodoped ITO NCs solution. Molar ratio percentage is defined as mol % = 100 × moles of F4TCNQ / (moles of F4TCNQ + moles of ITO NCs). $n_{reacted}$ is the number of F4TCNQ molecules that reacted with each ITO NC.